# Multicomponent Quantum Hall Ferromagnetism and Landau Level Crossing in Rhombohedral Trilayer Graphene

Y. Lee[1*], D. Tran[1], K. Myhro[1], J. Velasco Jr.[1], N. Gillgren[1], J.M. Poumirol[2], D. Smirnov[2], Y. Barlas[1]*, C. N. Lau[1*]

[1] Department of Physics and Astronomy, University of California, Riverside, Riverside, CA 91765
[2] National High Magnetic Field Laboratory, Tallahassee, FL 32310

**Using transport measurements, we investigate multicomponent quantum Hall (QH) ferromagnetism in dual-gated rhombohedral trilayer graphene (r-TLG), in which the real spin, orbital pseudospin and layer pseudospins of the lowest Landau level form spontaneous ordering. We observe intermediate quantum Hall plateaus, indicating a complete lifting of the degeneracy of the zeroth Landau level (LL) in the hole-doped regime. In charge neutral r-TLG, the orbital degeneracy is broken first, and the layer degeneracy is broken last and only the in presence of an interlayer potential $U_\perp$. In the phase space of $U_\perp$ and filling factor $\nu$, we observe an intriguing "hexagon" pattern, which is accounted for by a model based on crossings between symmetry-broken LLs.**

---

[*] Emails: kenosis101@gmail.com, yafisb@gmail.com; lau@physics.ucr.edu

In the quantum Hall (QH) regime, when the energies of two or more Landau levels (LLs) are brought to alignment, the spinor language is often used to describe the different degrees of freedom, such as layer and orbital pseudospins, due to their close analogy to the spins in a two-dimensional ferromagnet. When these LLs are less than completely full, competition between these degrees of freedom leads to formation of electronic states with spontaneous ordering of pseudospins, much like the spontaneous real spin alignment in a ferromagnet. For this reason, these symmetry-broken QH states are called QH ferromagnets, with real or pseudo-spin orderings that maybe easy-plane, *i.e.* akin to a XY Heisenberg magnet, or easy-axis, *i.e.* akin to an Ising ferromagnet. These QH ferromagnetic states provide a rich platform for investigation of the competition among different symmetries, as well as providing insight into the itinerant magnetism in standard magnets.

The recent emergence of two-dimensional (2D) graphene provides new playground for multicomponent QH ferromagnetic states and the associated phase transitions[1-6, 7-19, 20-28]. With the advent of high mobility samples that may be either suspended[29, 30] or supported on BN substrates[31, 32], and advanced device geometry such as dual-gates or split top gates[33-35], few-layer graphene provides QH systems with unusual symmetries and unprecedented tunability.

In particular, rhombohedral trilayer graphene is such a QH system with very flat bands near the charge neutrality point. Its LL energies are given by $E_N = \pm \frac{(2\hbar v_F eB)^{3/2}}{\gamma_1^2}\sqrt{N(N-1)(N-2)}$ where $N$ is an integer denoting the LL index, $e$ the electron charge, $v_F$~$10^6$ m/s the Fermi velocity of single layer graphene, $\gamma_1$~0.3 eV the interlayer hopping energy, and $h$ Planck's constant. The degeneracy between the $N$=0, 1 and 2 LLs, together with the spin and valley degrees of freedom, yield the 12-fold degeneracy of the lowest LL, and give rise to plateaus at filling factors $\nu$=±6, ±10, ±14… Interactions and/or single particle effects lift this 12-fold degeneracy, leading to incompressible QH ferromagnetic states at intermediate fillings[36, 37], with expected ordering of the real spin, the valley pseudospin and the orbital pseudospin. The order at which the degeneracy is broken reflects the underlying competing symmetries. Prior works have reported resolution of several symmetry-broken QH states[38-40], albeit only in single-gated samples where the interlayer potential $U_\perp$ and charge density $n$ are not independently controlled.

In this Letter, by using transport measurements on high mobility dual-gated r-TLG devices, we explore symmetry-broken LLs via careful control of $U_\perp$, $B$ and $n$. All integer plateaus of the zeroth LL in the hole-doped regime are resolved in high $B$. For charge balanced r-TLG sheets, even integer plateaus at filling factor ($\nu$=-2, -4 and -6) are resolved prior to that at odd integers ($\nu$=-1, -3 and -5), which are only resolved at finite $U_\perp$. We thus identify the QH states at intermediate filling factors as spin and pseudospin QH ferromagnets, which are filled in accordance to a "Hund's rule" of maximizing orbital pseudospin, followed by real spin and valley pseudospins. At constant $B$, we observe an intriguing "hexagon" patterns in $G(U_\perp,\nu)$ phase diagram, which arises from crossings between symmetry-broken LLs.

Dual-gated suspended devices are fabricated using a multi-level lithography technique [41, 42] (Fig. 1b). Two-terminal transport measurement is performed at 270mK using standard lock-in techniques in a He3 cryostat. We determine the contact resistance by subtracting a single resistance value from the entire data set, so that the conductance values of quantum Hall plateaus are properly quantized. The contact resistance thus obtained ranges from 0.2 to 1.2 kΩ. $U_\perp$ and $n$ are independently controlled via modulation of back gate voltage ($V_{bg}$) and top gate voltage ($V_{tg}$)[43]. Similar data are observed in two devices. Here we present data from a device with field effect mobility 42,000 cm$^2$/Vs.

High quality r-TLG devices are intrinsically insulating with a large interaction-induced gap, ~ 42 meV. This gapped insulating state is most likely a layer antiferromagnet with broken time reversal and spin rotation symmetries[44]. As $B$ increases from 0, the insulating state evolves smoothly into the $v$=0 QH state, which is most likely a canted antiferromagnetic phase[44], similar to that observed in bilayer graphene[10, 16-18]. Shubnikov-de Haas oscillations appear at $B$ as low as 0.2T, and conductance plateaus become quantized at $B$>3T. For devices with aspect ratio of unity, the two terminal conductance is given by $G = \sqrt{\sigma_{xx}^2 + \sigma_{xy}^2}$, where $\sigma_{xx}$ and $\sigma_{xy}$ are longitudinal and Hall conductivities. Thus $G$ is expected to be quantized at the QH plateaus where $\sigma_{xx}$=0, and display step-like behavior between plateaus[45]. Since our devices have aspect ratio slightly larger than 1 (width slightly larger than length), conductance between the plateaus is almost step-like, but with slightly non-monotonicity[45]; conductance quantization occurs at the "dip" regions of the $G(v)$ curves at a given filling factor. The strength of a given QH state can be assessed by whether a feature appears at said filling factor that moves with both $B$ and $n$, by the $B$ value at which it is first resolved, by deviation of $G$ from the expected value of $ve^2/h$ and by the width of the QH plateau.

The $v$=0 QH state is insulating with an in-plane anti-ferromagnetic order: the orientation of the in-plane is such that the spins in the top and bottom layers are in aligned opposite directions, however, due to the Zeeman splitting the in-plane spins get canted out of the xy-plane. The $v$=0 canted anti-ferromagnetic state is similar to the $v$=0 QH state observed in monolayer and bilayer graphene. This state spontaneously orders a particular combination of spin and valley degrees of freedom, thereby leaving two six-fold degenerate hole and electron LLs – a two-fold degeneracy associated with the spin/valley flavors and a three-fold degeneracy with the orbital pseudo-spin.

How are the remaining symmetries broken? Conventionally, for spin and valley independent SU(4) interactions, one expects a "Hund's rule" in which the triplet orbital degeneracy is the last to be broken[46, 47], i.e. $v$=-3 plateau is resolved first (Fig. 1e). This appears to be supported by data from singly-gated devices in this (Fig. 1c-d) and prior experiments[38]. As shown in Fig. 1c, which plots two terminal conductance $G(B, v)$ in units of $e^2/h$ *with top gate disconnected or grounded*, QH plateaus appear as vertical bands. As $B$ increases, QH plateaus at filling factors $v$=-5 and -3 are resolved first, followed by -1 and -2. This can be more clearly seen in the line traces $G(v)$: only the $v$=−3 plateau are fully resolved at $B$=4.5T, whereas additional plateaus at $v$=-2 and -1 are resolved at $B$=5.5T (Fig. 1d). These results are fully consistent with a prior work using singly-gated devices[38].

However, in the above measurement with only a single gate, $U_\perp$ is not controlled but scales with $n$. When we carefully control both $U_\perp$ and $n$, a qualitatively different picture emerges. Fig. 2a and 2b plots $G(B,v)$ at $U_\perp$=0 and -20 mV, respectively, and line traces at $B$=5T are shown in Fig. 2c and 2d. At $U_\perp$=-20 mV, the $v$=−3 (but not the $v$=−2) state is resolved (Fig. 2b and 2d), in an apparent agreement with the "Hund's rule", again qualitatively similar to data from singly-gated studies (Fig. 1c-d and ref. [38]). However, in the absence of interlayer bias, the plateaus at $v$=−6 are first resolved (as expected), followed by (unexpectedly) −4 and -2 that are fully resolved at $B$=5T; the odd integer plateaus $v$=−3 and -1 appear only as small shoulders even at $B$=8T. Thus, the exact sequence of plateaus depends strongly on $U_\perp$, thus the exact symmetries of QH states cannot be inferred from singly-gated devices. In particular, *for charge-balanced r-TLG, the orbital pseudospin is maximized first, i.e.* the triple orbital degeneracy is broken prior to that of spin and valley (Fig. 1f); this suggests that the conventional "Hund's rule" does not apply in this system. This indicates that either interactions are spin and valley dependent or that single particle hopping terms can influence the broken symmetry sequence (see discussion later).

To further investigate the plateaus' dependence on $U_\perp$ we measure $G(U_\perp, \nu)$ at constant $B$. The resolved QH plateaus appear as an array of bands centered at integer values of $\nu$, with a striking network of staggered "hexagons" (Fig. 3a-b). As $B$ increases, the sizes of these hexagons grow accordingly. $G$ is properly quantized at $\nu e^2/h$, except at certain critical $U_{\perp c}$ values that yield the horizontal "ridges" of the hexagons. For instance, at $\nu=-1$, $G$ is quantized at $1e^2/h$ except near $U_{\perp c} = 0$ mV (Fig. 3c, green curve); at $\nu=-2$ and $\nu=-4$ states, quantization is lost at $U_{\perp c} \sim \pm 18$ mV (red curve) and $U_{\perp c} \sim 35$ mV (blue curve), respectively, and the corresponding $U_{\perp c}$ values are indicated by ■ and ▲. Consequently, a given plateau is resolved (unresolved) if $G(\nu)$ is taken at $U_\perp \neq U_{\perp c}$ ($U_\perp = U_{\perp c}$), e.g. the line traces in Fig. 2c and 2d are effectively taken along the red and green arrows in Fig. 3b, respectively.

Our experimental results demonstrate the presence of QH ferromagnetic states in r-TLG, and enable us to determine the symmetries of the states at intermediate filling factors. One combination of the spin-valley degrees of freedom is lifted first, leading to the layer antiferromagnetic state at $\nu=0$. This is followed by the breaking of the orbital degeneracy and the appearance of the even integer states at $\nu=-2$ and -4. Finally, in the presence of finite $U_\perp$ that breaks the inversion symmetry, the remaining spin-valley symmetries are broken and the odd integer states are resolved. Schematic of the symmetries of the QH states in the lowest LL is shown in Fig. 4a.

Within the QH ferromagnetism, the hexagon patterns can be naturally accounted for by a model of crossings between LLs[48-50], whose energies depend on both $U_\perp$ and $B$. In the lowest LL of the two-band model, only the A (B) sublattice of the top (bottom) layer are relevant for low-energy considerations. We thus ignore the contribution of the middle layer, and treat layer, valley and sublattice indices as equivalent. Hence, energies of LLs that are partially localized to the top (bottom) layer increase (decrease) with increasing $U_\perp$; these two sets of LLs cross whenever the difference in LL energies are compensated by the externally applied interlayer potential. At a given $\nu$, $G$ is quantized properly *except* at the crossing points. Using LL spectra similar to that depicted in Fig. 4a, we model the density of states of each LL as a Lorentzian and calculate the total density of states of the system as a function of $n$ and $U_\perp$. The simulation result reproduces the observed "hexagon" patterns (Fig. 4b), confirming the presence of multiple LL crossings driven by $U_\perp$ in the QH regime.

In principle, we can determine the LL gaps $\Delta$ from $U_{\perp c}$ at the LL crossing points, where the differences in LL energies are compensated by electrostatic energy. However, here $U_\perp$ is the externally imposed potential bias, and will be heavily screened[51-53] even in the QH regime. Thus one expects that $\Delta(B) = U_\perp^S(B) \ll U_\perp$, where $U_\perp^S$ is the *screened* interlayer potential. Extracting the exact magnitude of $\Delta$ from the crossing points is non-trivial and will be the focus of future studies. Nevertheless, we do not expect screening to significantly alter the functional dependence of $U_\perp(B)$. Hence insight into the nature of the broken-symmetry QH states in the lowest LL can be obtained by examining the dependence of $U_{\perp c}$ on $B$.

To this end, we plot $U_{\perp c}(B)$ for crossings observed at $\nu=-2$ and -4 in Fig. 4c. Interestingly, $U_{\perp c}(B)$ is linear in $B$ for $\nu=-2$ state, but markedly sub-linear for the $\nu=-4$ state. Thus the data in Fig. 4d suggest that the LL gap of the $\nu=-2$ state scales linearly with $B$, whereas that of the $\nu=-4$ state is sub-linear in $B$. The different scaling behaviors in $B$ for the $\nu=-2$ and $\nu=-$

4 gaps suggest different mechanisms of gap generation. In particular, the remote hopping term $\gamma_4$ in r-TLG, which is the interlayer hopping energy between stacked-unstacked sublattices, may also lead to splitting of the orbital degeneracy of the zeroth LL. This effect can be captured in an effective two-band model[54], evaluated in the perturbation theory $H_{\gamma_4} = \frac{2v_4 v_F}{\gamma_1}\begin{pmatrix} \pi^+\pi & 0 \\ 0 & \pi\pi^+ \end{pmatrix}$, where $v_4 = \sqrt{3}a\gamma_4/2\hbar$, $a$=0.246 nm is the lattice constant, and $\pi=\pm p_x+ip_y$. In the presence of $B$, $H_{\gamma_4}$ leads to a splitting of the $N$=0, 1 and 2 LL orbitals. Such splitting of the orbital pseudospin has an energy gap $\Delta_{\gamma_4}$ that scales linearly with $B$, and leads to QH plateaus at $\nu=\pm 2$, as observed experimentally. In fact, ignoring other remote hopping parameters, reasonable agreement between experimental data and LL spectrum can be obtained by using $\gamma_4 \sim 0.1\, \gamma_1$, though this crude estimate may be modified by other hopping terms and by the non-zero potential at the middle layer. On the other hand, the sub-linear behavior for the $\nu$=-4 state suggests an origin of electronic interactions, which lifts the spin-valley degeneracy and is expected to scale with $\sqrt{B}$.

The contrasting behavior of the in the scaling gaps of the $\nu$=-2 and $\nu$=-4 states suggest that both the single particle remote hopping terms and electron-electron interactions in the zeroth LL must be included to account for the broken symmetries in the zeroth LL. We also find that the addition of the remote hopping terms can significantly influence the Hund's rules determining the Hall plateau sequence of the broken symmetry states. Taken together, our data suggest that the $\nu$=-2 and -4 QH states are orbital pseudospin polarized canted antiferromagnetic states, whereas the $\nu$=-1, -3 and -5 states, resolved only in the presence of finite $U_\perp$, are layer/spin polarized. Further theoretical and experimental studies, such as those using samples with even higher quality, or graphene/hexagonal boron nitride heterostructures[31] for measurements of LL gaps and crossings for large ranges of magnetic field, electric field and charge densities, are needed to understand the mechanism of gap generation in the orbital pseudospin indices, and to help determine the precise values of remote hopping parameters in few-layer graphene.

**Acknowledgements.**
We thank R. Cote for discussions. This work is supported by DOE BES Division under grant no. ER 46940-DE-SC0010597. CNL acknowledges support by CONSEPT center at UCR. Part of this work was performed at NHMFL that is supported by NSF/DMR-0654118, the State of Florida, and DOE.

**Supporting Information.**
Device fabrication and characterization, Landau fan diagram and simulations of Landau level crossings at constant magnetic field.

# References

1. Novoselov, K. S.; Geim, A. K.; Morozov, S. V.; Jiang, D.; Katsnelson, M. I.; Grigorieva, I. V.; Dubonos, S. V.; Firsov, A. A. *Nature* **2005,** 438, (7065), 197-200.
2. Zhang, Y. B.; Tan, Y. W.; Stormer, H. L.; Kim, P. *Nature* **2005,** 438, (7065), 201-204.
3. Das Sarma, S.; Adam, S.; Hwang, E. H.; Rossi, E. *Rev. Mod. Phys.* **2011,** 83, 407.
4. Fuhrer, M. S.; Lau, C. N.; MacDonald, A. H. *MRS Bull.* **2010,** 35, (4), 289-295.
5. Peres, N. M. R.; Guinea, F.; Castro Neto, A. H. *Ann. Phys.* **2006,** 321, (7), 1559-1567.
6. Castro Neto, A. H.; Guinea, F.; Peres, N. M. R.; Novoselov, K. S.; Geim, A. K. *Rev. Mod. Phys.* **2009,** 81, (1), 109-162.
7. Jung, J.; Zhang, F.; MacDonald, A. H. *Phys. Rev. B* **2011,** 83, (11), 115408.
8. Min, H.; Borghi, G.; Polini, M.; MacDonald, A. H. *Phys. Rev. B* **2008,** 77, (4), 041407.
9. Nagaosa, N.; Sinova, J.; Onoda, S.; MacDonald, A. H.; Ong, N. P. *Rev. Mod. Phys.* **2010,** 82, (2), 1539-1592.
10. Zhang, F.; Jung, J.; Fiete, G. A.; Niu, Q. A.; MacDonald, A. H. *Phys. Rev. Lett.* **2011,** 106, (15), 156801.
11. Zhang, F.; MacDonald, A. H. *Phys. Rev. Lett.* **2012,** 108, 186804.
12. Zhang, F.; Min, H.; MacDonald, A. H. *Phys. Rev. B* **2012,** 86, 155128.
13. Zhang, F.; Min, H.; Polini, M.; MacDonald, A. H. *Phys. Rev. B* **2010,** 81, (4), 041402 (R).
14. Nandkishore, R.; Levitov, L. *Phys. Rev. B* **2010,** 82, (11), 115124.
15. Nandkishore, R.; Levitov, L. *Phys. Rev. Lett.* **2011,** 107, (9), 097402.
16. Velasco, J.; Jing, L.; Bao, W.; Lee, Y.; Kratz, P.; Aji, V.; Bockrath, M.; Lau, C. N.; Varma, C.; Stillwell, R.; Smirnov, D.; Zhang, F.; Jung, J.; MacDonald, A. H. *Nature Nanotechnol.* **2012,** 7, 156-160.
17. Maher, P.; Dean, C. R.; Young, A. F.; Taniguchi, T.; Watanabe, K.; Shepard, K. L.; Hone, J.; Kim, P. *Nat. Phys.* **2013,** 9, (3), 154-158.
18. Kharitonov, M. *Phys. Rev. Lett.* **2012,** 109, 046803.
19. Checkelsky, J. G.; Li, L.; Ong, N. P. *Phys. Rev. Lett.* **2008,** 100, (20), 206801.
20. Abanin, D. A.; Parameswaran, S. A.; Sondhi, S. L. *Phys. Rev. Lett.* **2009,** 103, (7), 076802.
21. Cote, R.; Jobidon, J. F.; Fertig, H. A. *Phys. Rev. B* **2008,** 78, 085309-1-13.
22. Cote, R.; Luo, W. C.; Petrov, B.; Barlas, Y.; MacDonald, A. H. *Phys. Rev. B* **2010,** 82, (24), 245307.
23. Doucot, B.; Goerbig, M. O.; Lederer, P.; Moessner, R. *Phys. Rev. Lett.* **2008,** 78, (19), 195327.
24. Lu, C.-K.; Herbut, I. F. *Phys. Rev. Lett.* **2012,** 108, (26), 266402.
25. Shibata, N.; Nomura, K. *J. Phys. Soc. Jpn.* **2009,** 78, (10), 104708.
26. Toke, C.; Lammert, P. E.; Crespi, V. H.; Jain, J. K. *Phys. Rev. B* **2006,** 74, (23), 235417.
27. Yang, K.; Sarma, S. D.; MacDonald, A. H. *Phys. Rev. B* **2006,** 74, (7), 075423.
28. Young, A. F.; Dean, C. R.; Wang, L.; Ren, H.; Cadden-Zimansky, P.; Watanabe, K.; Taniguchi, T.; Hone, J.; Shepard, K. L.; Kim, P. *Nat. Phys.* **2012,** 8, (7), 550-556.
29. Bolotin, K. I.; Sikes, K. J.; Jiang, Z.; Klima, M.; Fudenberg, G.; Hone, J.; Kim, P.; Stormer, H. L. *Sol. State Commun.* **2008,** 146, (9-10), 351-355.
30. Du, X.; Skachko, I.; Barker, A.; Andrei, E. Y. *Nat. Nanotechnol.* **2008,** 3, (8), 491-495.

31. Dean, C. R.; Young, A. F.; Meric, I.; Lee, C.; Wang, L.; Sorgenfrei, S.; Watanabe, K.; Taniguchi, T.; Kim, P.; Shepard, K. L.; Hone, J. *Nat. Nanotechnol.* **2010,** 5, (10), 722-726.
32. Taychatanapat, T.; Watanabe, K.; Taniguchi, T.; Jarillo-Herrero, P. *Nat. Phys.* **2011,** 7, 621.
33. Weitz, R. T.; Allen, M. T.; Feldman, B. E.; Martin, J.; Yacoby, A. *Science* **2010,** 330, 812-816.
34. Maher, P.; Wang, L.; Gao, Y.; Forsythe, C.; Taniguchi, T.; Watanabe, K.; Abanin, D.; Papić, Z.; Cadden-Zimansky, P.; Hone, J.; Kim, P.; Dean, C. R. *Science* **2014,** 345, (6192), 61-64.
35. Ozyilmaz, B.; Jarillo-Herrero, P.; Efetov, D.; Abanin, D. A.; Levitov, L. S.; Kim, P. *Phys. Rev. Lett.* **2007,** 99, 166804.
36. Cote, R.; Rondeau, M.; Gagnon, A.-M.; Barlas, Y. *Phys. Rev. B* **2012,** 86, (12), 125422.
37. Barlas, Y.; Cote, R.; Rondeau, M. *Phys. Rev. Lett.* **2012,** 109, (12), 126804.
38. Elferen, H. J. v.; Veligura, A.; Tombros, N.; Kurganova, E. V.; Wees, B. J. v.; Maan, J. C.; Zeitler, U. *Phys. Rev. B* **2013,** 88, 121302 (R).
39. Bao, W. Z.; Zhao, Z.; Zhang, H.; Liu, G.; Kratz, P.; Jing, L.; Velasco, J.; Smirnov, D.; Lau, C. N. *Phys. Rev. Lett.* **2010,** 105, (24), 246601.
40. Kumar, A.; Escoffier, W.; Poumirol, J. M.; Faugeras, C.; Arovas, D. P.; Fogler, M. M.; Guinea, F.; Roche, S.; Goiran, M.; Raquet, B. *Phys. Rev. Lett.* **2011,** 107, (12), 126806.
41. Liu, G.; Velasco, J.; Bao, W. Z.; Lau, C. N. *Appl. Phys. Lett.* **2008,** 92, (20), 203103.
42. Velasco, J.; Liu, G.; Bao, W. Z.; Lau, C. N. *New J. Phys.* **2009,** 11, 095008.
43. McCann, E.; Koshino, M. *Rep. Prog. Phys.* **2013,** 76, (5), 056503.
44. Lee, Y.; Tran, D.; Myhro, K.; Jr., J. V.; Gilgren, N.; Barlas, Y.; Poumirol, J. M.; Smirnov, D.; Guinea, F.; Lau, C. N. *Nat. Commun.* **2014**, in press (an earlier vesion available at arXiv:1402.6413).
45. Abanin, D. A.; Levitov, L. S. *Phys. Rev. B* **2008,** 78, (3), 035416.
46. Zhang, F.; Tilahun, D.; MacDonald, A. H. *Phys. Rev. B* **2012,** 85, (16), 165139.
47. Barlas, Y.; Cote, R.; Nomura, K.; MacDonald, A. H. *Phys. Rev. Lett.* **2008,** 101, (9), 097601.
48. Zhang, X.; Faulhaber, D.; Jiang, H. *Phys. Rev. Lett.* **2005,** 95, (21), 216801.
49. Sanchez-Yamagishi, J. D.; Taychatanapat, T.; Watanabe, K.; Taniguchi, T.; Yacoby, A.; Jarillo-Herrero, P. *Phys. Rev. Lett.* **2012,** 108, (7), 076601.
50. Lee, K.; Fallahazad, B.; Xue, J.; Taniguchi, T.; Watanabe, K.; Tutuc, E. *Science* **2014,** 345, 58-61.
51. van Gelderen, R.; Olsen, R.; Smith, C. M. *preprint* **2013**, arXiv:1304.5501.
52. Koshino, M. *Physical Review B* **2010,** 81, (12), 125304.
53. Min, H.; Hwang, E. H.; Das Sarma, S. *Phys. Rev. B* **2012,** 86, (8), 081402.
54. Zhang, F.; Sahu, B.; Min, H. K.; MacDonald, A. H. *Phys. Rev. B* **2010,** 82, (3), 035409.

Fig. 1. (a-b). Band structure and SEM image of TLG device. (c). $G(B,v)$ of a r-TLG device with only back gate engaged. (d). Line traces $G(v)$ at $B=4.5T$ and $5.5T$, respectively. (e-f). Schematics of orders of symmetry breaking in r-TLG in the QH regime. Fig 1e shows the filling sequence anticipated from valley and spin independent Coulomb interactions whereas Fig 1f indicates that the filling sequence is modified with the addition of remote hopping effects (see text for details).

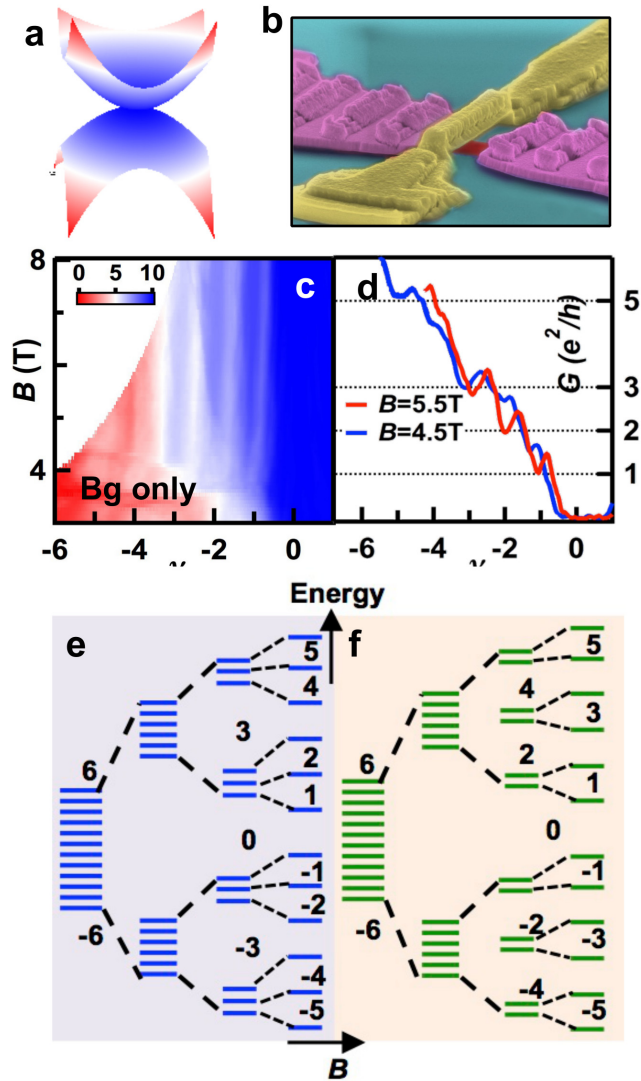

Fig. 2. (a)(c). $G(B,\nu)$ in units of $e^2/h$ at $U_\perp=0$, and line traces $G(\nu)$ at $B=5$T. (b)(d). Similar data at $U_\perp=-20$ mV.

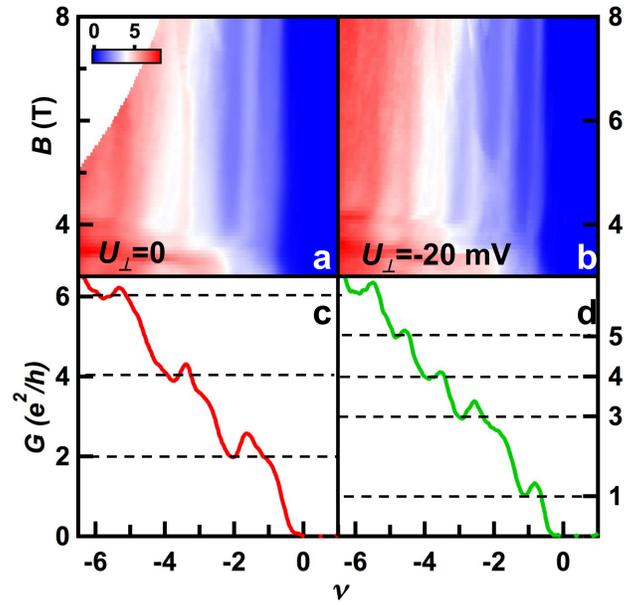

Fig. 3. (a-b). $G(U_\perp, \nu)$ in units of $e^2/h$ at $B$=7T and 5T, respectively. The arrows indicate line traces along which Fig. 2c and 2d would be taken. (c). Line traces $G(U_\perp)$ at $B$=5T and $\nu$=-1, -2 and -4. The triangle and squares mark $U_{\perp c}$ values at which $G$ is not quantized.

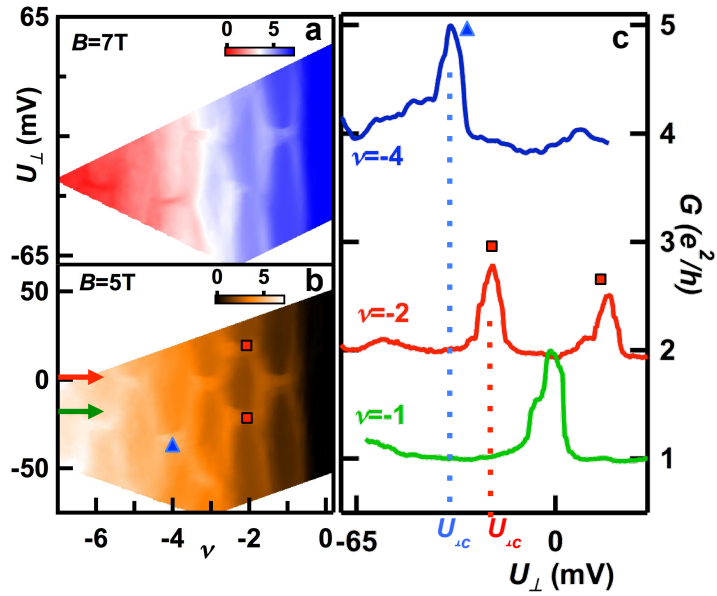

Fig. 4. (a). Schematic diagram of LL evolution with $U_\perp$ and the resultant QH states in the hole-doped regime. Colored numbers, ± and arrows indicate orbital, valley and spin indices. (b). Simulated total density of states vs. $U_\perp$ and $\nu$. Color scale: blue (low), red (high). (c). Measured $U_{\perp c}(B)$ for $\nu$=-2 and -4 states, respectively. The dotted lines are guides to the eye.

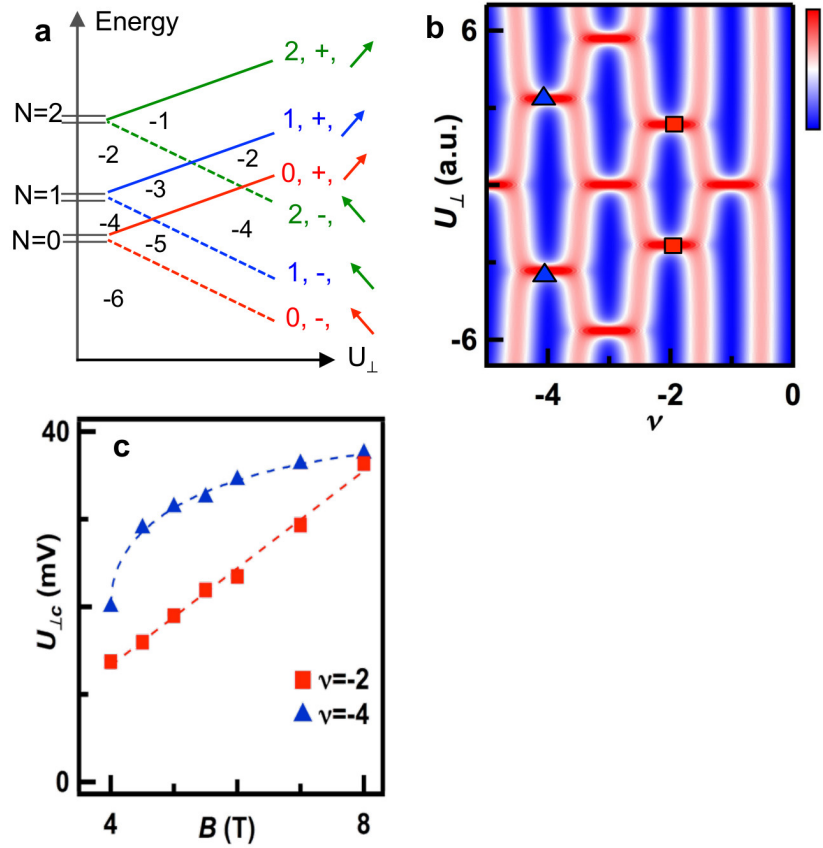

For Table of Contents Only

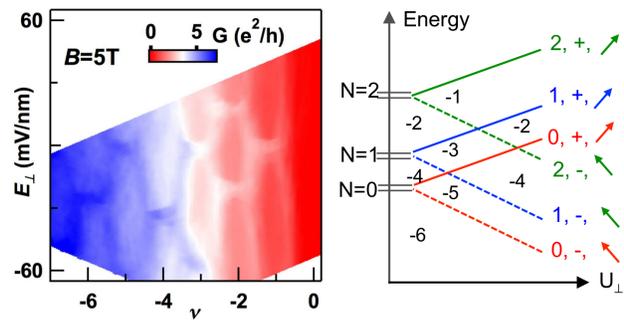